\documentclass{article}

\usepackage{PRIMEarxiv}

\usepackage{placeins}
\usepackage{amsmath}
\usepackage{algorithm}
\usepackage{algpseudocode}
\usepackage[utf8]{inputenc} % allow utf-8 input
\usepackage[T1]{fontenc}    % use 8-bit T1 fonts
\usepackage{hyperref}       % hyperlinks
\usepackage{url}            % simple URL typesetting
\usepackage{booktabs}       % professional-quality tables
\usepackage{amsfonts}       % blackboard math symbols
\usepackage{nicefrac}       % compact symbols for 1/2, etc.
\usepackage{microtype}      % microtypography
\usepackage{lipsum}
\usepackage{fancyhdr}       % header
\usepackage{graphicx}       % graphics
\usepackage{hyperref}       % links
\usepackage{listings}
\usepackage{xcolor}
\graphicspath{{media/}}     % organize your images and other figures under media/ folder

\title{An Algorithmic Approach to Line Construction in Existing Transit Networks
}

\author{
  Zezhi Deng, Ruoxing Yang \\
  Georgetown University \\
  Washington, D. C\\
  \texttt{\{zd127, ry216\}@georgetown.edu} \\
}

\begin{document}
\maketitle

\begin{abstract}
Transit networks often have existing infrastructure that cannot be modified when designing new lines for the network. This paper provides an algorithm to generate a line within a transit network without changing any existing lines or connections between stations. Additionally, a method of analyzing the efficiency of a transit line and network is provided, and used within the algorithm presented. An analysis of the effects of different parameters and objectives on the location of a new line is performed. We find that under most cases, a new line generated improves the overall efficiency of the network, while under certain circumstances, an unsuitable combination of pathfinding algorithm and efficiency evaluation method or an increase in construction and maintenance cost can cause the algorithm to create a less efficient network.
\end{abstract}

% keywords can be removed

\section{Introduction}
Public transit is crucial in keeping cities functional and moving people to their destinations efficiently, and efficient public transit networks allow for the movement of large amounts of people through urban areas, while reducing emissions and congestion from car journeys  \cite{importance}. However, the pressure put on these networks is growing as a result of the ever-expanding populations in the urban areas of cities. As a result, the public transit networks of many cities are constantly under expansion, with additional lines being constructed and connected to the existing network.

Given this need, numerous works have been published on the Transit Network Design Problem (TNDP), where a solution to TNDP aims to generate a set of lines on a given graph of transit nodes or stations within a city. Durán-Micco and Vansteenwegen provide a survey on existing TNDP studies \cite{survey}. A solution to the TNDP consists of a set of transit lines on the network and a set of frequencies associated with the lines. For the purpose of this paper, we only focus on the generation of transit lines, as the generation of frequencies are outside of this scope and is another well-analyzed problem. However, TNDP algorithms are designed to be run on a transit network with easy-to-modify connections, such as a bus network, as the algorithms aim to generate a set of lines from scratch, without any existing lines in the network.

In this paper, we produce an algorithm to add a line to a transit network without modifying any existing stations, connections, or lines. Many major urban areas already have some form of an established metro system, or other forms of permanent or semi-permanent transit infrastructure such as trams, overground light rail, or bus rapid transit networks. Given this, establishing new connections and rerouting existing lines can be expensive and in many cases infeasible, as existing transit route infrastructure, such as tunnels and tram tracks, are expensive to modify, and in many cases cannot be modified feasibly. Therefore, solutions to TNDP may not be useful for many public transit projects, and different algorithms and methods of analysis need to be used in order to determine the suitable location of new lines in the transit network.

This paper is structured as follows: Section 2 outlines the existing work on efficiency analysis of transit networks, TNDP algorithms, and other works related to public transit efficiency improvement. The motivations behind this study is also discussed in this section. In Section 3, we provide definitions and notation, followed by network evaluation methods and the Line Addition Algorithm. In Section 4 we provide a case study based on the Washington Metro network. A conclusion of the study and future work is provided in Section 5.

\section{Background and Motivation}

Laporte and Pascoal propose a path based algorithm that generates specific locations for metro lines and stations along a general corridor predetermined by city planners \cite{corridor}. Owais et al. present a model to construct new metro networks, in a grid or ring-radial design, within cities based on the non-demand criterion of minimizing passenger transfers \cite{non-demand}. Cancela et al. note that most solutions to TNDP employ one of three strategies: heuristics, metaheuristics, or mathematical programming \cite{math-programming} . Durán-Micco and Vansteenwegen observe that in most TNDP studies, an objective function is comprised of a combination of user and operator costs, where the main user cost is in-vehicle travel time plus a fixed penalty for transfers, and operator costs is a combination of construction costs and operating costs, typically proportional to the length of the lines in the network \cite{survey}. In this paper, we create an objective function with similar considerations.

We focus on methods inspired by the Route Generation Algorithm (RGA) of Baaj and Mahmassani \cite{rga} and the Pair Insertion Algorithm (PIA) of Mauttone and Urquhart \cite{pia} in our algorithm for the Line Addition Problem. RGA generates a set of initial route skeletons by picking the highest demand station pairs, and expands each line skeleton into a full line by inserting suitable stations into the line. Multiple criteria are used in determining whether a station is suitable for insertion, including the amount of demand from the station that remains unfulfilled, the insertion cost of the station into the line, and the length and circuity of the line following the insertion \cite{rga}. PIA repeatedly selects the node pair with the highest demand between them that is not covered by a line, and chooses the lower cost option between inserting it into an existing line or generating a new line from the shortest path between the two nodes \cite{pia}. Both PIA and RGA rely on an origin-destination (O-D) matrix that contains information on the demand between all stations.

Many TNDP algorithms require a graph that contains all feasible connections between stations (i.e. if stations A and B are in adjacent regions, then there exists an edge between them on the graph). For example, PIA takes a graph G where vertices are stations or zones in the city, and edges between vertices exist if the two zones are adjacent or a connection can be logically created \cite{pia}. In the case of metro networks, doing so is a lot less feasible as metro stops don't have to be in directly adjacent areas for a feasible connection to exist between them. For example, on the WMATA red line, every station on the west side from Shady Grove to Farragut North can feasibly be connected to every station on the east side from Fort Totten to Glenmont. Such considerations greatly complicate the generation of the input graph for TNDP problems. Therefore, a method of creating new connections without the input of a graph with possible connections is needed. 

We define the Line Addition Problem as the generation of a new transit line within an existing transit network that generates the greatest improvement on overall efficiency, while not changing any existing lines or connections, and present the Line Addition Algorithm as a potential solution. The Line Addition Algorithm takes the elements of line expansion from RGA and pair insertion from PIA and adapts them to focus on the generation of a single line instead of an entire network. As the addition of a line has factorial complexity with regard to the number of stations, approximate methods have to be used to determine an efficient way of line insertion, especially when dealing with networks with a large amount of existing infrastructure and demand, such as the metro networks of cities like London, Paris, New York, or Beijing. Additionally, in the generation of lines between station pairs, we reference the work of Laporte and Pascoal in locating a metro line through a given corridor \cite{corridor}. Doing so allow connections to be formed between the new line and the existing metro network without the need of a pre-generated connectivity graph between stations.

Another problem related to the Line Addition Problem that has been studied is the extension of existing transit lines. Matisziw, Murray, and Kim propose an algorithm that extends transit lines based on spatial optimization, where increased demand is maximized while minimizing added distance on the line \cite{routeextension}. The extension of transit lines into areas without previous coverage is important in improving the efficiency of transit networks, as it allows for more people without previous access to the transit network to access the network. Thus although line extension is not a focus of this study, it plays an important role in designing new transit lines and can be used in conjunction with the algorithms proposed in this paper.

One important set of factors we consider in the generation of new lines is the efficiency of existing lines, paths through a network, and networks as a whole. Literature on the efficiency evaluation of public transit networks is rich. Brons, Nijkamp, Pels, and Rietveld's study comparing evaluation methods contains an overview of popular methods, detailing how frontier analysis is commonly used to evaluate the technical efficiency (TE) of a public transit network \cite{efficiency-evaluation}. Our algorithm addressing the Line Addition Problem iteratively applies an efficiency evaluation function to existing lines and intermediary products. Thus, we chose a simpler, runtime-efficient approach to evaluate transit efficiency where travel time minimization is prioritized when judging a network. Broadly, we define network efficiency as the sum of individual path efficiencies, where individual path efficiencies are defined as transit time spent on the path as a proportion to direct geographic distance from the origin to the destination. 

Finally, we note that demand to add a new line to an existing network occurs on an infrequent basis. The most recent addition to the Washington Metrorail system (WMATA) was the silver line extension, a project which began in 2014 and opened in 2022. Recognizing this, we demonstrate that the value of this project is not solely in algorithmically automating the line-generation process, but rather in constructing a systematic framework for analyzing the effects of weighing different user preferences on line generation. In the case study section, we provide an overview of the effects of adjusting different parameters on generating a new line for the Washington Metrorail.

\section{Algorithm}

\subsection{Definitions and Notation}
We model public transit networks using a graph $G = (V, E)$, where each vertex $v_i \in V$ represents a station in the network, and each edge $e_{ij} \in E$ represents an existing connection between $v_i$ and $v_j$ in the network, such as a metro or rail line. The distance of an existing connection $l_{ij}$ is the length of the edge $e_{ij}$, and $l_{ij}^*$ represents the distance as-the-crow-flies between two stations. Demand between stations is represented as an origin-destination matrix $D = \{d_{ij} | i, j \in \{1, \ldots, n\}\}$, where each $d_{ij}$ denotes the number of travelers from station $i$ to station $j$. A line $r_i$ in a network is a sequence of edges $\left(e_{ij}, e_{jk}, \ldots\right)$ such that the destination of the previous edge in the sequence is the origin of the next edge. We denote the set of all lines as $R$. A path through the network is a sequence of edges $\left(e_{ij}, e_{jk}, \ldots\right)$ and a set of lines $\{r_i, r_j, \ldots\}$ that the path travels through. It is important to note that though we define lines as a sequence of edges, they can be treated as sets, as in a line with no loops, there is one unique way to order the edges such that they form a line, therefore information on the order of edges is redundant. As such, we apply set operations such as union and difference on lines in the algorithms presented.

\subsection{Network Evaluation Methods}

\paragraph{Construction Cost}
The cost of constructing a new line is linear to the distance covered by the connections. When considering the cost of construction, an average of global cost data or an average of local or state data can be used. The cost of construction per unit length as derived from empirical data is c, and the cost of constructing a line is:
\begin{equation*}
\text{cost(r)} = c * \text{r.length}
\end{equation*}

\paragraph{Pathfinding in the Network}
When considering how a transit user travels through the network between two stations, different models can be used to generate the path that the user takes, which we refer to as the best path function. Different pathfinding algorithms can be used for the best path function, including shortest path functions such as A*, or transfer minimizing functions such as PathPlanning \cite{pathplanning}. The function path($v_i$, $v_j$) will be used in subsequent algorithms to represent the best path function. 

\paragraph{Path Complexity}
The cost of a single passenger moving through one path within the network is represented as the complexity of the path. Complexity is measured in arbitrary units and is calculated as the sum of total distance covered by each connection, amount of station stops, and amount of line transfers. Each variable is weighted by a specified constant parameter. We define the weight of the number of transfers as $w_t$, the weight of the number of stations on the path as $w_s$, and the weight of the total distance of the path as $w_d$. The amount of station stops is reduced by 2 to discount the origin and destination stations. The path complexity function can be defined as
\begin{equation*}
\text{pathComplexity(path($v_i, v_j$))} = \text{path.transfers} * w_t+ \text{(path.stations - 2)} * w_s + \text{path.distance} * w_d
\end{equation*}

\paragraph{Path Efficiency}
The efficiency of a path is defined as the ratio of the complexity of that path to the as-the-crow-flies geographical distance between the origin and destination stations of the path. The geographical distance is scaled by adjustment weight $w_a$. Note that a higher efficiency number indicates that the path between a station pair is more complex in relation to a most optimal direct connection, and therefore signifies a worse efficiency. The ratio is raised to a power to penalize lower efficiency paths further. In this paper, we raise the efficiency to a power of $4$ to scale efficiency similarly to demand, but other powers and functions can be used depending on the circumstances. Since the arbitrary units that measure complexity and the geographical units used to measure distance both model usage cost, they can be compared to produce a usable ratio factor used in other calculations. Path efficiency is modeled as
\begin{equation*}
\text{pathEfficiency(path($v_i, v_j$))} = (\text{(pathComplexity(path($v_i, v_j$))}/(l_{ij}^* * w_a))^4
\end{equation*}

\paragraph{Network Efficiency}
The total efficiency of the entire transit network is calculated as the product of the sum of the path complexities between all possible station pairs, weighted by the demand of the path, multiplied by the total construction cost of all lines within the network. The regression scheme is customizable to account for the logarithmic nature of complexity variation. A lower returned value represents a better efficiency measure. Line efficiency is evaluated by considering a line as an isolated network and conducting network efficiency analysis.
\begin{equation*}
    \text{totalEfficiency} = \sum_{v_i, v_i \in V} \text{pathEfficiency(path($v_i, v_j$))} * d_{ij}
\end{equation*}

Depending on the regression scheme, the function networkEfficiency can return different values. The schemes that can be used are linear-linear, linear-log, log-linear, and log-log. For example, a linear-log regression scheme would result in
\begin{equation*}
    \text{networkEfficiency} = \left(\sum_{r_i \in R} \text{cost($r_i$)}\right) * \ln\left({\text{totalEfficiency}}\right)
\end{equation*}
and a log-linear regression scheme would result in
\begin{equation*}
    \text{networkEfficiency} = \ln\left(\sum_{r_i \in R} \text{cost($r_i$)}\right) * \text{totalEfficiency}
\end{equation*}

\paragraph{Assignment Method}
We use a naïve all-or-nothing assignment method when planning an individual’s path through the network \cite{modelbook}. That is, when considering the path from station A to station B in path efficiency calculations, effects of congestion on path choice and travel time are not considered, and individuals take the most efficient path regardless of the choice of others. We believe this approach suffices for basic analysis of public transit networks, as the travel time between stations on a vehicle is not impacted by the amount of people on the vehicle, and the impact of congestion on the wait times at stations is not significant compared to other factors impacting travel time.

\subsection{Line Addition Algorithm}
We use a greedy algorithm in selecting the stations to create the new line. The basic concept of this algorithm is to identify the station pairs in the network that have the worst efficiency, and prioritize forming lines between those stations. Since station pairs with bad efficiency can be spread across the network, they often cannot be joined with one line that satisfies length and circuity constraints. Therefore, a list of lines under construction, called lineCandidates, is maintained, where each line candidate is a line that satisfies circuity and length constraints, which are constructed from some low efficiency station pairs chosen by the algorithm. The algorithm iteratively identifies the least efficient station pair, and tries to add it to the most suitable line in lineCandidates, and if no suitable line exists, it constructs a new line between the stations. Lines in lineCandidates can be joined if the ends of the lines overlap, and the resulting line satisfies all the line constraints.

While constructing lines, there are certain constraints that have to be satisfied. $L_{min}$ represents the minimum length of a new line, and $L_{max}$ represents the max length. Lines in the network should not become too circuitous; the travel distance between two stations on the line should not be too long in relation to the direct distance. We define the circuity factor $\rho$ as the ratio between the travel distance between any two stations on the line and the direct distance between the two stations. As such, $\rho_{ij} = \text{path($vi, vj$)}/l_{ij}^*$, and a line satisfies the circuity constraint if for all station pairs $v_i, v_j$, $\rho_{ij} < \rho_{max}$. To simplify calculations, $\rho$ can alternatively be defined as the ratio of travel distance on the line between the two terminal stations and the direct distance between the terminal stations, as defined in \cite{pia}. Doing so, however, could create situations where certain segments of the line are overly circuitous.

\begin{algorithm}
\begin{algorithmic}
\caption{Line Addition Algorithm}
\Require $G = (V, E), D = \{d_{ij}\}$, targetEfficiency
\Ensure A line r in the network G
\State $\text{lineCandidates} \gets \emptyset$
\State $E \gets \text{createEfficiencyMatrix(D)}$
\While {$!\text{targetEfficiencySatisfied()}$ and $!E.\text{empty}$}
\State $\epsilon_{ij} \gets $ the worst efficiency station pair in $E$
\State $r_m', r_n', r \gets $ null
\If{$\exists r_m \in R$ such that $v_i \in r_m$}
    \State $r_m' \gets $ addToLine($v_j, r_m$)
    \State lineCandidates $\gets$ lineCandidates $\cup  \{r_m'\} - \{ r_m\}$
    \For {each $r_k \in$ lineCandidates}
        \State $r_k' \gets $ joinLine($r_k, r_m'$)
        \If{$r_k' \neq$ null}
            \State lineCandidates $\gets$ lineCandidates $\cup  \{r_k'\}$
        \EndIf
    \EndFor
\ElsIf{$\exists r_n \in R$ such that $v_j \in r_n$}
    \State $r_n' \gets $ addToLine($v_i, r_n$)
    \State lineCandidates $\gets$ lineCandidates $\cup  \{r_n'\} - \{ r_n\}$
    \For {each $r_k \in$ lineCandidates}
        \State $r_k' \gets $ joinLine($r_k, r_n'$)
        \If{$r_k' \neq$ null}
            \State lineCandidates $\gets$ lineCandidates $\cup  \{r_k'\}$
        \EndIf
    \EndFor
\EndIf
\If{$r_m', r_n' =$ null}
    \State $r \gets$ constructLine($v_i, v_j$)
    \State lineCandidates $\gets$ lineCandidates $\cup  \{r\}$
\EndIf
\State remove from $E$ all node pairs directly covered by $r, r_m', r_n'$
\State removeSubsetLines()
\State updateEfficiencies()
\EndWhile
\State \Return findBestLine()
\end{algorithmic}
\end{algorithm}

Two methods can be used when considering the termination condition of the algorithm. One is to iterate through every station pair in the efficiency matrix, then find the most efficient line. However, doing so increases the cost of computation, as the efficiency matrix will have many station pairs with low demand between them, which do not impact line generation significantly. Therefore, we return the first line that satisfies target efficiency and minimum length constraints. A function targetEfficiencySatisfied() can be defined, that checks if there is a line within lineCandidates that satisfies line length and circuity constraints, and has an efficiency lower than targetEfficiency. Doing so can reduce the computational complexity, but will also not guarantee termination of the algorithm or the selection of the most suitable line, as the algorithm might terminate before an efficient line is found, or the target efficiency is too low for any line to achieve. When choosing a target efficiency for the termination condition, an arbitrary value can be chosen based on test runs. Alternatively, a target efficiency can be chosen based on the average efficiency of the lines already in the network, or the best or worst efficiency of a line in the network.

\paragraph{Efficiency and Demand Matrix}
One important consideration is that adding a connection between two stations doesn’t just benefit those two stations, as other paths that pass through those two stations could take advantage of the new connection, so taking this into account is important when determining which station pair to choose in line construction. Therefore a modified demand matrix is produced by the createEfficiencyMatrix($E$) function. This function generates a set of paths between all possible station pairs. Using this set, it first calculates efficiencies between each station pair. Then, it produces a modified demand matrix. The demand for each station pair $v_i$, $v_j$  is augmented by accounting for the demand between all station pairs $a$, $b$ where $a$ and $b$ are the origin and destination stations of any path that includes the path $v_i$, $v_j$ as a sub-path. The additional demand has to be weighted, as the farther a traveler has to travel from $a$ to $v_i$ or from $v_j$ to $b$, the less important a new connection between $v_i$ and $v_j$ will be to a traveler going from $a$ to $b$. Therefore, we find the sum of the distance between $a$ and $v_i$ and the distance between $v_j$ and $b$,  and define calculateDemandWeight($p_i$, $p_j$) as $1/(1 +$ distance * $w)$, where $w$ is an adjustment weight. In the case study section, $w$ is set to 10. The path then has its efficiency evaluated, and the efficiency is multiplied with the modified demand.

\begin{algorithm}
    \begin{algorithmic}
        \caption{createEfficiencyMatrix($D$)}
        \Require Demand Matrix $D$
        \Ensure Efficiency Matrix $E$
        \State modifiedDemand $\gets$ \{\}
        \State paths $\gets$ \{path($v_i, v_j$)| $v_i, v_j \in V, d_{ij} \in D$\}
        \For{each path($v_i, v_j$) $\in$ paths}
            \State additionalDemand $\gets 0$
            \For {each path($v_m, v_n$) $\in$ paths}
                \If{$p_i \subset p_j$}
                    \State additionalDemand $\gets$ additionalDemand + $d_{mn} *$ calculateDemandWeight($p_i, p_j$)
                \EndIf
            \EndFor
            \State modifiedDemand$_{ij} \gets d_{ij} +$ additionalDemand
        \EndFor
        \State $E \gets \{\epsilon_{ij} | \epsilon_{ij} \gets \text{pathEfficiency}(v_i, v_j) * \text{modifiedDemand}_{ij} \}$
        \State \Return $E$
    \end{algorithmic}
\end{algorithm}

As lines are created in lineCandidates, the efficiency of the connection between stations will change, since new connections and lines in lineCandidates can shorten the path between stations. Therefore, there is no longer a need to consider a new connection between a station pair that previously had a low efficiency if a new connection is established that improves the efficiency. updateEfficiencies() performs this update by selecting the worst efficiency station pair $(v_i, v_j)$, recalculating path($v_i, v_j$) by including all lines in lineCandidates as well as the existing network, and checking if it is still the worst efficiency station pair. If it is, then the function can be terminated. If it is not, then select the new worst efficiency station pair and repeat this operation. This ensures that the station pair selected by the next iteration of the algorithm is indeed the worst efficiency station pair after line generation in lineCandidates, while minimizing the number of path($v_i, v_j$) operations needed.

\begin{algorithm}
    \begin{algorithmic}
        \caption{updateEfficiencies()}
        \Repeat
        \State $\epsilon_{ij} \gets $ the worst efficiency station pair in $E$
        \State $\epsilon_{ij} \gets \text{pathEfficiency}(v_i, v_j) * \text{modifiedDemand}_{ij}$
        \Until{$\epsilon_{ij}$ = the worst efficiency station pair in $E$}
    \end{algorithmic}
\end{algorithm}

\paragraph{Construct Line Operation}
When we consider constructing a new line between two stations $v_i$ and $v_j$, a line consisting of only the edge $e_{ij}$ could suffice, but doing so would miss the opportunity to connect the edge to other stations that are geographically close to $e_{ij}$ to create transfer opportunities. PIA and RGA solve this issue by having a pre-existing graph of feasible connections between stations, and finding the shortest path between the two stations on the graph. However, as outlined previously, such an approach is not suitable for the Line Addition Problem. Therefore, we take an approach similar to the corridor-based line generation of Laporte and Pascoal, where suitable stations (those with high demand to and from the stations already in the line) are considered as transfer stations to the candidate line under construction \cite{corridor}. It is important to limit the number of stations under consideration for addition into the new line, so we generate a corridor between $v_i$ and $v_j$, and only consider the stations that lie within the corridor. Then we use a recursive approach in selecting the station $v_k$ with the highest demand to $v_i$ and $v_j$, then consider all stations in the corridor from $v_i$ to $v_k$, and from $v_k$ to $v_j$. For the purpose of this algorithm, we consider the corridor between $v_i$ and $v_j$ as a rhombus, with $v_i$ and $v_j$ as the endpoints of the longer diagonal, and the shorter diagonal is a proportion of the distance between $v_i$ and $v_j$. Other corridor designs can be considered. See Figure \ref{fig:constructLine} in Appendix A for a diagram of the process.

\begin{algorithm}
\begin{algorithmic}
    \caption{constructLine($v_i$, $v_j$)}
    \Require $v_i, v_j \in V$
    \Ensure a line $r$
    \State $r \gets$ null
    \State $v_k \gets$ null
    \State maxDemand $\gets -\infty$
    \State define targetArea as corridor between $v_i$ and $v_j$
    \State stationCandidates $\gets \{ v_i | v_i \in \text{targetArea}\}$
    \If{stationCandidates $=$ null}
        \State $r = \{e_{ij}\}$
    \Else
        \For{each station $v_m \in$ stationCandidates}
        \State demand $\gets d_{im} + d_{mj}$
        \If {demand $>$ maxDemand}
            \State maxDemand $\gets$ demand
            \State $v_k \gets v_m$
        \EndIf
        \EndFor
        \State $r =$ constructLine($v_i$, $v_k$) $\cup$ constructLine($v_k$, $v_j$)
    \EndIf
    \State \Return $r$
\end{algorithmic}
\end{algorithm}

Additionally, since the construct line operation could produce a line that entirely overlaps with another line, we can define the removeSubsetLine() operation to remove any line from lineCandidates that is entirely contained within any other line in lineCandidates or the existing network. Since lines are defined as a set of edges, the operation removes any line that is a subset of another line.

\paragraph{Add to Line Operation}
If one of the line candidates already contains a station from the chosen station pair, we could add that station pair to the existing line, and extend the end of the line towards the other station using the construct line operation. In the following operation, $v_i$ is the station already contained in line $r_i$. To find the most suitable location for the $v_j$ in $r_i$, we take inspiration from the candidates function in PIA \cite{pia}. If there are $p$ stations in $r_i$, then there are $p + 1$ valid positions for inserting $v_j$, and the algorithm finds the insertion position with the lowest cost. If the resulting line from the insertion satisfies all the line constraints, then it will be returned, otherwise no candidate line has been found, and the procedure returns null. When inserting a station into a position $p$, constructLine($v_p - 1$, $v_j$) and constructLine($v_j$, $v_p$) is called to connect $v_j$ to the existing stations. If $v_j$ is being inserted at the start of the line, only constructLine($v_j$, $v_p$) is called, and if it is being inserted at the end of the line, only constructLine($v_p - 1$, $v_j$) is called.

\begin{algorithm}
    \begin{algorithmic}
        \caption{addToLine($v_j, r_i$)}
        \Require $v_j \in V$, $r_i \in$ lineCandidates
        \Ensure a modified line $r'_i$
        \State $r'_i \gets$ null, $r_{temp} \gets$ null
        \State minCost $\gets \infty$ 
        \For {$p \gets 1, \dots, |r_i| + 1$}
            \State $r_{temp} \gets$ insert $v_j$ into $r_i$ at position $p$
            \If{cost($r_{temp}) <$ minCost and line constraints are satisfied}
                \State $r'_i \gets r_{temp}$
            \EndIf
        \EndFor
        \State \Return $r'_i$
    \end{algorithmic}
\end{algorithm}

\paragraph{Join Line Operation}
When constructing line candidates, it is possible that two or more lines are constructed such that the end of one line entirely overlaps with the start of another line. In this case, if circuity and max length constraints are satisfied, the two lines can be joined to form a longer line, which can reduce construction cost and transfer complexity. Since we have established that edges in a line is unique and lines can be treated as a set of edges, we can use the set union operation to form the new combined line.

\begin{algorithm}
    \begin{algorithmic}
        \caption{joinLine($r_i$, $r_j$)}
        \Require $r_i, r_j \in$ lineCandidates
        \Ensure a line $r_k$ containing all stations from $r_i$ and $r_j$
        \State $r_k \gets$ null
        \If{the last $n$ stations of $r_i$ = the first $n$ stations of $r_j$}
            \State $r_k \gets r_i \cup r_j$ 
        \ElsIf{the first $n$ stations of $r_i$ = the last $n$ stations of $r_j$}
            \State $r_k \gets r_i \cup r_j$ 
        \EndIf
        \State \Return $r_k$
    \end{algorithmic}
\end{algorithm}

\subsection{Notes}
The addToLine procedure can be replaced by the candidates procedure in PIA. This changes how station pairs are inserted into existing lines in lineCandidates, where a station pair can be inserted into a line when neither station is already a part of that line. Doing so increases the computational complexity of the algorithm, as there are ${p \choose 2}$ positions for each station pair in a line, and every line has to be considered in the insertion. Through constructing lines between station pairs using constructLine(), most lines in lineCandidates would have already considered inserting stations that satisfies circuity constraints, so iterating through every line in lineCandidates when finding a suitable insertion location for a station pair is largely redundant.

When creating the graph for the algorithm, planners can consider whether a station is suitable to be turned into a transfer station, and only include those that are in the graph $G$. This way all lines generated using the Line Addition Algorithm will be connected to the existing network at suitable transfer stations. All stations still have to be included in demand calculation, since the modified demand between two stations is calculated based on the total number of travelers travelling through the station pair, not just the trips originating and ending at the two stations.

\section{Case Study: Washington Metro}
In this section, we demonstrate the Line Addition Algorithm in action by applying it to the Washington Metro network. As noted throughout the algorithm, several key parameters impact the design of a new line. These parameters can be categorized into two groups: efficiency evaluation parameters and line construction parameters. Efficiency evaluation parameters adjust the relative weights of line properties during efficiency evaluation. Line construction parameters specify necessary constraints when building new connections and lines. See tables ~\ref{tab:efficiency} and ~\ref{tab:construction} in Appendix B. We use demand data provided from PlanItMetro, the online planning blog from WMATA \cite{website:planitmetro}, and cost data from the Eno Center for Transportation \cite{website:costdata}.

The following section presents new lines generated by the Line Addition Algorithm for the Washington Metro network based on varying parameter configurations. These sample results demonstrate the effects on algorithm behavior of modifying individual efficiency and construction parameters. We provide sets of demonstrative parameters derived from observing results from our test runs of the algorithm. We intend for and encourage the user to explore algorithmic behavior by providing their own parameter configurations. We also compare total network efficiency before and after the addition of the new line. Note that a lower efficiency value is considered better efficiency, thus a negative percentage improvement indicates a positive improvement on the network. The A* path-finding algorithm was used for the best path function. All results were generated using version 1.0.0 of the Line Addition Algorithm.

\paragraph{Notes on Minimum Length} Minimum length and target efficiency are the two key parameters that govern the termination of the algorithm. For all experiments run on the Washington Metro network, we selected 8.2 as the baseline minimum length constraint. This value was obtained by taking the approximate square root of 68.3, the geographic area in square miles of Washington D.C.

Most existing lines consist of three main sections:

Section 1: A series of stations exclusive to the line extending from the origin station into the main metropolitan area.

Section 2: A series of stations that are typically shared with other lines and are located within the main metropolitan area.

Section 3: A series of stations exclusive to the line extending from the main metropolitan area towards the destination station.

Since the Line Addition Algorithm does not add new stations to the network, generated lines will not have any stations exclusive to the line. They thus will not possess typical section 1 and section 3 components. We choose the approximate cross-section size of the main metropolitan area as the minimum length constraint to allow the algorithm to consider a line that only consists of a section 2 component as a valid line. Generated lines are shorter than typical existing lines because they do not possess section 1/section 3 extensions beyond the metropolitan area.

\subsection{Case 1: Baseline Parameters}

Line Construction Configuration:

\begin{lstlisting}
p-max:1.5
max-length:35
min-length:8.2
corridor-height:0.5
demand-adjustment-weight:10
target-efficiency:150
\end{lstlisting}

Efficiency Evaluation Configuration:

\begin{lstlisting}
cost-mode:state-average-DC
line-regression:log-log
regression:log-log
transfer-weight:-1
station-weight:-0.75
distance-weight:1
adjustment-weight:1
\end{lstlisting}

Result:

Line:
\begin{lstlisting}
eastern market -> union station
union station -> judiciary square
judiciary square -> gallery place-chinatown
gallery place-chinatown -> metro center
metro center -> u street-cardozo
u street-cardozo -> columbia heights
columbia heights -> silver spring
\end{lstlisting}

Metrics:
\begin{lstlisting}
Total algorithm runtime in minutes: 15
Old network efficiency: 217.37074433876535
New network efficiency: 216.57277364458218
Improvement: -0.7979706941831637
Percentage improvement: -0.3671012383062699%
\end{lstlisting}

Map: See Figure \ref{fig:line1} in Appendix C.

\subsection{Case 2: Increasing Minimum Length}

In this example, we increase the minimum length constraint to 10. We relax target efficiency to 175 to accommodate the tighter constraint. 

Line Construction Configuration:

\begin{lstlisting}
p-max:1.5
max-length:35
min-length:10
corridor-height:0.5
demand-adjustment-weight:10
target-efficiency:175
\end{lstlisting}

Efficiency Evaluation Configuration:

\begin{lstlisting}
cost-mode:state-average-DC
line-regression:log-log
regression:log-log
transfer-weight:-1
station-weight:-0.75
distance-weight:1
adjustment-weight:1
\end{lstlisting}

Result:

Line:
\begin{lstlisting}
farragut west -> foggy bottom
foggy bottom -> rosslyn
rosslyn -> east falls church
east falls church -> west falls church
west falls church -> tysons corner
tysons corner -> greensboro
greensboro -> vienna
\end{lstlisting}

Metrics:
\begin{lstlisting}
Total algorithm runtime in minutes: 7
Old network efficiency: 217.37074433876535
New network efficiency: 218.90776077119398
Improvement: 1.5370164324286293
Percentage improvement: 0.7070944331097465%
\end{lstlisting}

Map:See Figure \ref{fig:line2} in Appendix C.

\subsection{Case 3: Increasing Corridor Height}

In this example, we increase the size of the line construction corridor by increasing the corridor height to 1.0. This allows the algorithm to generate new connections in a larger area. 

Line Construction Configuration:

\begin{lstlisting}
p-max:1.5
max-length:35
min-length:8.2
corridor-height:1.0
demand-adjustment-weight:10
target-efficiency:150
\end{lstlisting}

Efficiency Evaluation Configuration:

\begin{lstlisting}
cost-mode:state-average-DC
line-regression:log-log
regression:log-log
transfer-weight:-1
station-weight:-0.75
distance-weight:1
adjustment-weight:1
\end{lstlisting}

Result:

Line:
\begin{lstlisting}
pentagon -> farragut west
farragut west -> farragut north
farragut north -> dupont circle
dupont circle -> woodley park-zoo
woodley park-zoo -> cleveland park
cleveland park -> van ness-udc
van ness-udc -> silver spring
\end{lstlisting}

Metrics:
\begin{lstlisting}
Total algorithm runtime in minutes: 3
Old network efficiency: 217.37074433876535
New network efficiency: 217.87379478553714
Improvement: 0.5030504467717947
Percentage improvement: 0.2314250927842464%
\end{lstlisting}

Map: See Figure \ref{fig:line3} in Appendix C.

\subsection{Case 4: Prioritizing Minimal Transfers}

In this example, we increase the weight adjustment of a line transfer from -1 to -0.5 in efficiency evaluation to study the effect of prioritizing minimal transfers during travel. Note that all baseline weights are 1.0, so a -0.5 weight adjustment results in a final weight of 0.5.

Line Construction Configuration:

\begin{lstlisting}
p-max:1.5
max-length:35
min-length:8.2
corridor-height:0.5
demand-adjustment-weight:10
target-efficiency:175
\end{lstlisting}

Efficiency Evaluation Configuration:

\begin{lstlisting}
cost-mode:state-average-DC
line-regression:log-log
regression:log-log
transfer-weight:-0.5
station-weight:-0.75
distance-weight:1
adjustment-weight:1
\end{lstlisting}

Result:

Line:
\begin{lstlisting}
pentagon -> farragut west
farragut west -> farragut north
farragut north -> columbia heights
columbia heights -> georgia avenue-petworth
georgia avenue-petworth -> silver spring
\end{lstlisting}

Metrics:
\begin{lstlisting}
Total algorithm runtime in minutes: 2
Old network efficiency: 218.9208053587901
New network efficiency: 220.26122765368416
Improvement: 1.3404222948940685
Percentage improvement: 0.6122863894535951%
\end{lstlisting}

Map: See Figure \ref{fig:line4} in Appendix C.

\subsection{Case 5: Relaxing Circuity Constraint}

In this example, we relax the circuitry constraint pMax from 1.5 to 2.0 to allow for curvier lines. 

Line Construction Configuration:

\begin{lstlisting}
p-max:2.0
max-length:35
min-length:8.2
corridor-height:0.5
demand-adjustment-weight:10
target-efficiency:150
\end{lstlisting}

Efficiency Evaluation Configuration:

\begin{lstlisting}
cost-mode:state-average-DC
line-regression:log-log
regression:log-log
transfer-weight:-1
station-weight:-0.75
distance-weight:1
adjustment-weight:1
\end{lstlisting}

Result:

Line:
\begin{lstlisting}
silver spring -> georgia avenue-petworth
georgia avenue-petworth -> columbia heights
columbia heights -> farragut north
farragut north -> farragut west
farragut west -> mcpherson square
mcpherson square -> metro center
metro center -> gallery place-chinatown
gallery place-chinatown -> judiciary square
judiciary square -> union station
union station -> capitol south
capitol south -> navy yard
\end{lstlisting}

Metrics:
\begin{lstlisting}
Total algorithm runtime in minutes: 2
Old network efficiency: 217.37074433876535
New network efficiency: 215.09252871024452
Improvement: -2.278215628520826
Percentage improvement: -1.0480783122177195%
\end{lstlisting}

Map: See Figure \ref{fig:line5} in Appendix C.

\section{Conclusion}
We propose an algorithm to generate a new line within a transit network without modifying any existing connections, with the aim of increasing the overall efficiency of the network. Additionally, we define a set of efficiency evaluation methods to be used in identifying the station pairs with a need of a new connection between them. We find that under certain cases, the algorithm generates a new line within the network that can improve the overall efficiency of the network. Under other constraints, the overall efficiency of the network can be lowered. This is because additional lines in a network can increase the average amount of transfers of paths through the network, and depending on how transfers are weighted in complexity calculation, extra transfers can decrease the efficiency of a path. Thus, to generate optimal results, we recognize that the pathfinding algorithm should match the weights in complexity calculation. For example, if transfers are weighted heavily in path complexity, then a transfer-minimizing pathfinding function should be used for path($v_i, v_j$). Additionally, as cost scales linearly with the distance of the transit line, an increase in cost of a network by adding a line can also decrease the overall efficiency of the network.

While we acknowledge that the design of new transit lines happens on a relatively infrequent basis, we believe that an algorithm that procedurally generates a line in an existing network based on demand and efficiencies between stations is valuable for urban planners to study the effects of prioritizing different considerations on the location and design of the new line.

The focus of this algorithm is on the determination of the path of a new metro line through an existing network, and therefore does not focus on the location of new metro stations along the line or the generation of specific connections. Numerous methods have been proposed to solve this problem, including the aforementioned path-based algorithms from Laporte and Pascoal, and parcel-level modeling of demand from Furth and Mekuria \cite{corridor} \cite{parcel}. Transit line extension algorithms, such as the algorithm proposed by Matisziw, Murray, and Kim, can be used to extend the new line into areas without existing metro connections \cite{routeextension}.

Further study can be performed on the effects of congestion on line efficiency. Though we think the effect of congestion on travel time through public transit networks is minimal compared to that of road networks, it is still possible that transit users choose to avoid stations and lines that are overcrowded. An additional consideration would be the avoidance of Braess’s Paradox, where under certain conditions adding capacity to a network could make everybody worse off \cite{modelbook}. 

\section{Availability}
Our implementation of the Line Addition Algorithm is available as free and open-source software (subject to the MIT license) and can be accessed from https://github.com/Davidrxyang/raines-2024-line-addition-problem.

\section*{Acknowledgments}
Funding for this project was granted by the Center for Research and Fellowships of Georgetown University, through the Lisa J. Raines Fellowship. The proposal for this project was supported by Professor Mark Maloof of Georgetown University. We would like to thank them, and other members of the Georgetown University community, who have made this paper possible.

%Bibliography
\bibliographystyle{unsrt}  
\bibliography{references}  

\newpage
\appendix

\section{Construct Line Operation}
\begin{figure}[h]
    \centering
    \includegraphics[width=1\linewidth]{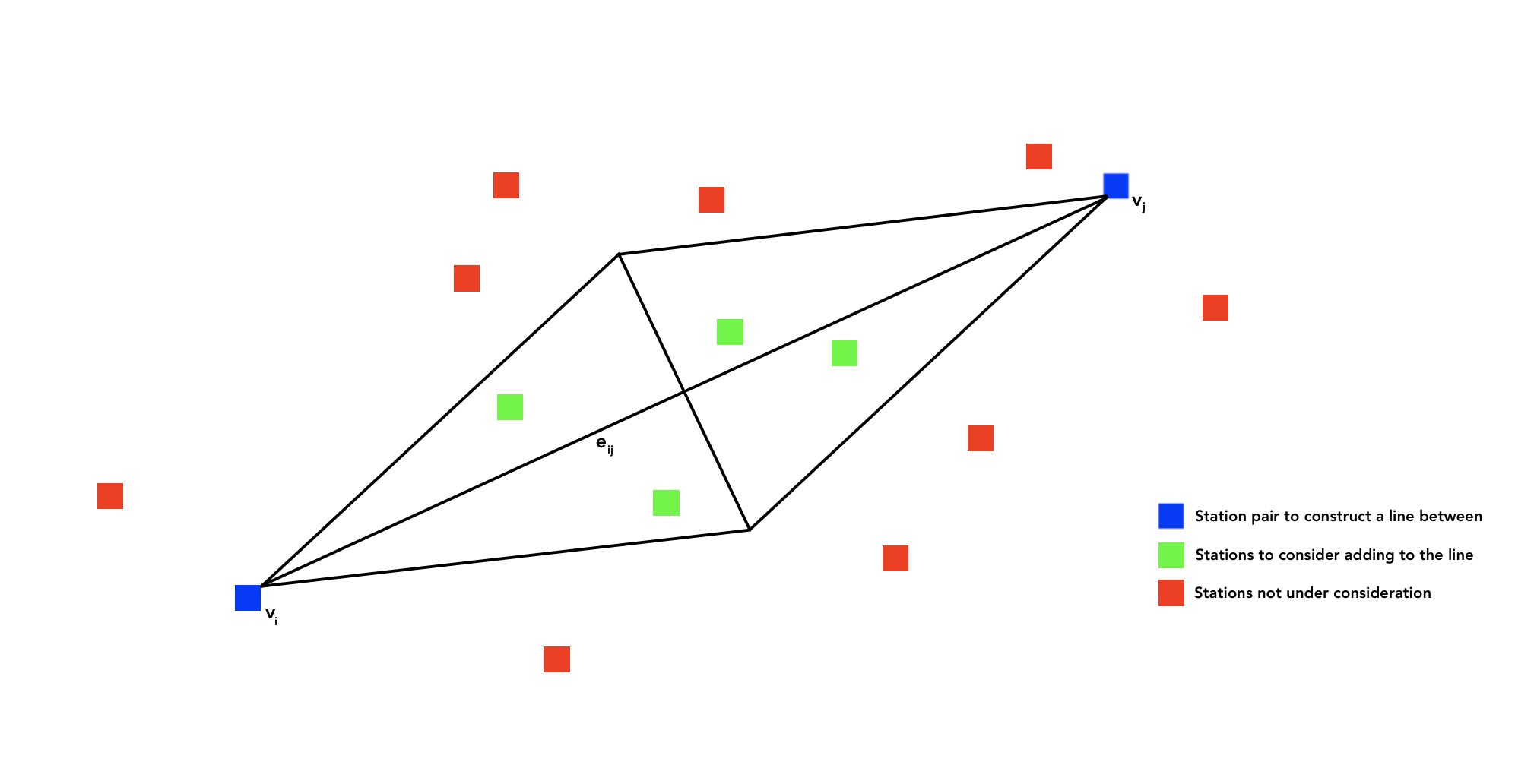}
    \caption{The process of selecting stations in operation constructLine. The rhomboidal area between the station pair represents targetArea}
    \label{fig:constructLine}
\end{figure}

\newpage
\section{Line Addition Algorithm Parameters}
Efficiency evaluation parameters adjust the relative weights of line properties during efficiency evaluation. Line construction parameters specify necessary constraints when building new connections and lines. 

\begin{table}[ht]
 \caption{Efficiency Evaluation Parameters}
  \centering
  \begin{tabular}{l|p{70mm}}
    Parameter     & Definition     \\
    \midrule
    $w_t$ & Transfer weight, the significance of the number of line transfers when evaluating the complexity of a path. \\
    \hline
    $w_s$ & Station weight, the significance of the number of station stops when evaluating the complexity of a path. \\
    \hline
    $w_d$ & Distance weight, the significance of the travel distance when evaluating the complexity of a path \\
    \hline
    $w_a$ & Adjustment weight, the significance of geographical distance when adjusting complexity by distance when evaluating the efficiency of a path \\
    \hline
    regression-mode & Mathematic functional form specification for cost-efficiency relationship. \\
    \hline
    cost-mode & cost evaluation scheme. \\
    \hline
    $c$ & Additional demand coefficient, the significance of additional covered demand when generating a path demand value for the modified demand table.  \\
    \hline
    targetEfficiency & Desired efficiency value, acts as a termination threshold. \\
    \bottomrule
  \end{tabular}
  \label{tab:efficiency}
\end{table}

\begin{table}[ht]
 \caption{Line Construction Parameters}
  \centering
  \begin{tabular}{l|p{70mm}}
    Parameter     & Definition     \\
    \midrule
    $L_{min}$ & Minimum line length, a lower bound on the length of a constructed line for it to be considered valid. \\
    \hline
    $L_{max}$ & Maximum line length, an upper bound on the length of a constructed line for it to be considered valid. \\
    \hline
    $\rho_{max}$ & The directness of a line, defined as the ratio of the travel distance between any two stations on a line and the direct geographical distance between the two stations.  \\
    \hline
    targetArea & Geographical area between two stations containing all valid stations to be potentially included in the new line. \\
    \bottomrule
  \end{tabular}
  \label{tab:construction}
\end{table}

\newpage
\section{Case Study Maps}

\begin{figure}[H]
    \centering
    \includegraphics[width=0.5\linewidth]{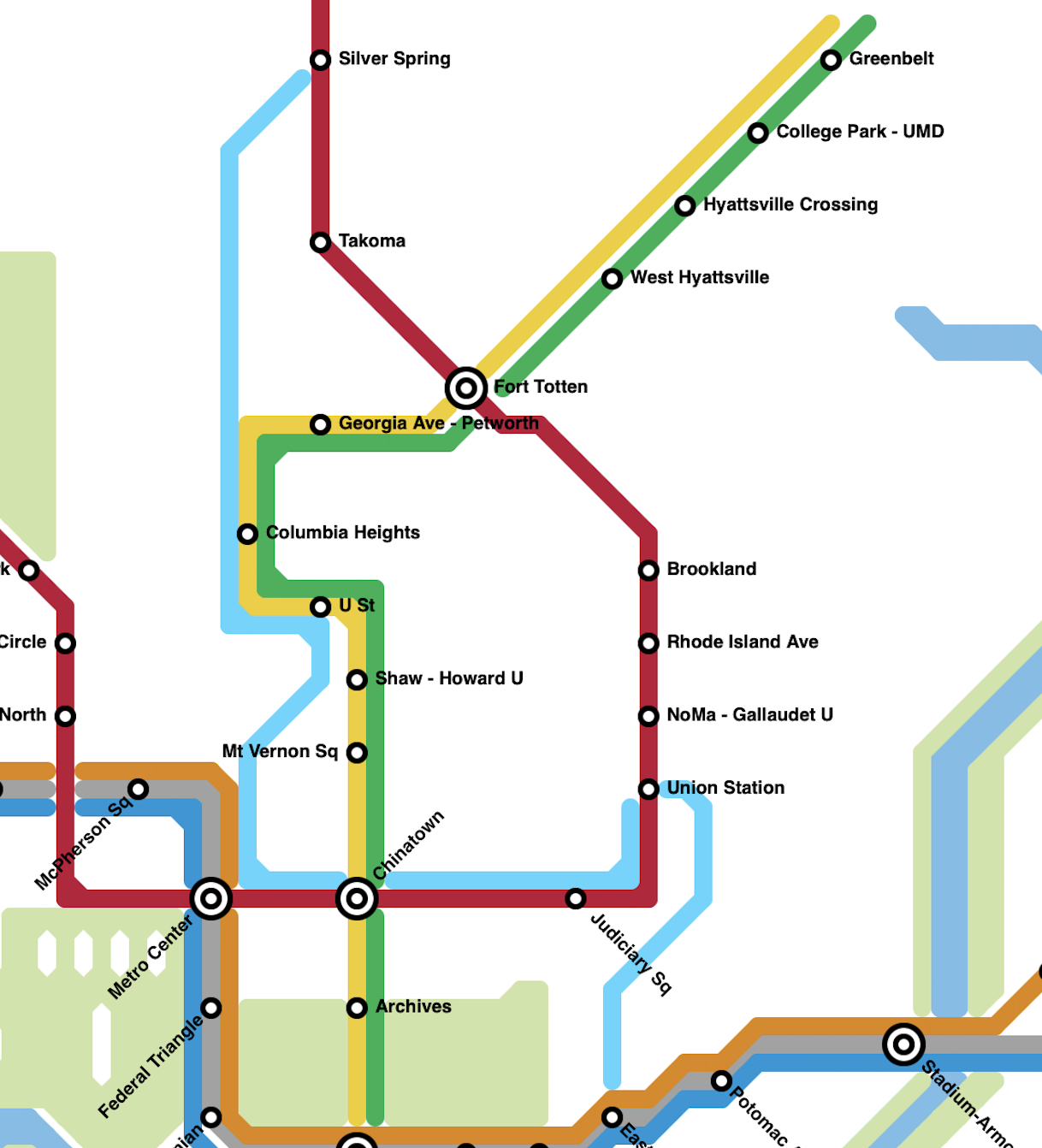}
    \caption{Map of WMATA metro with line in case 1 added in cyan}
    \label{fig:line1}
\end{figure}

\begin{figure}[H]
    \centering
    \includegraphics[width=0.5\linewidth]{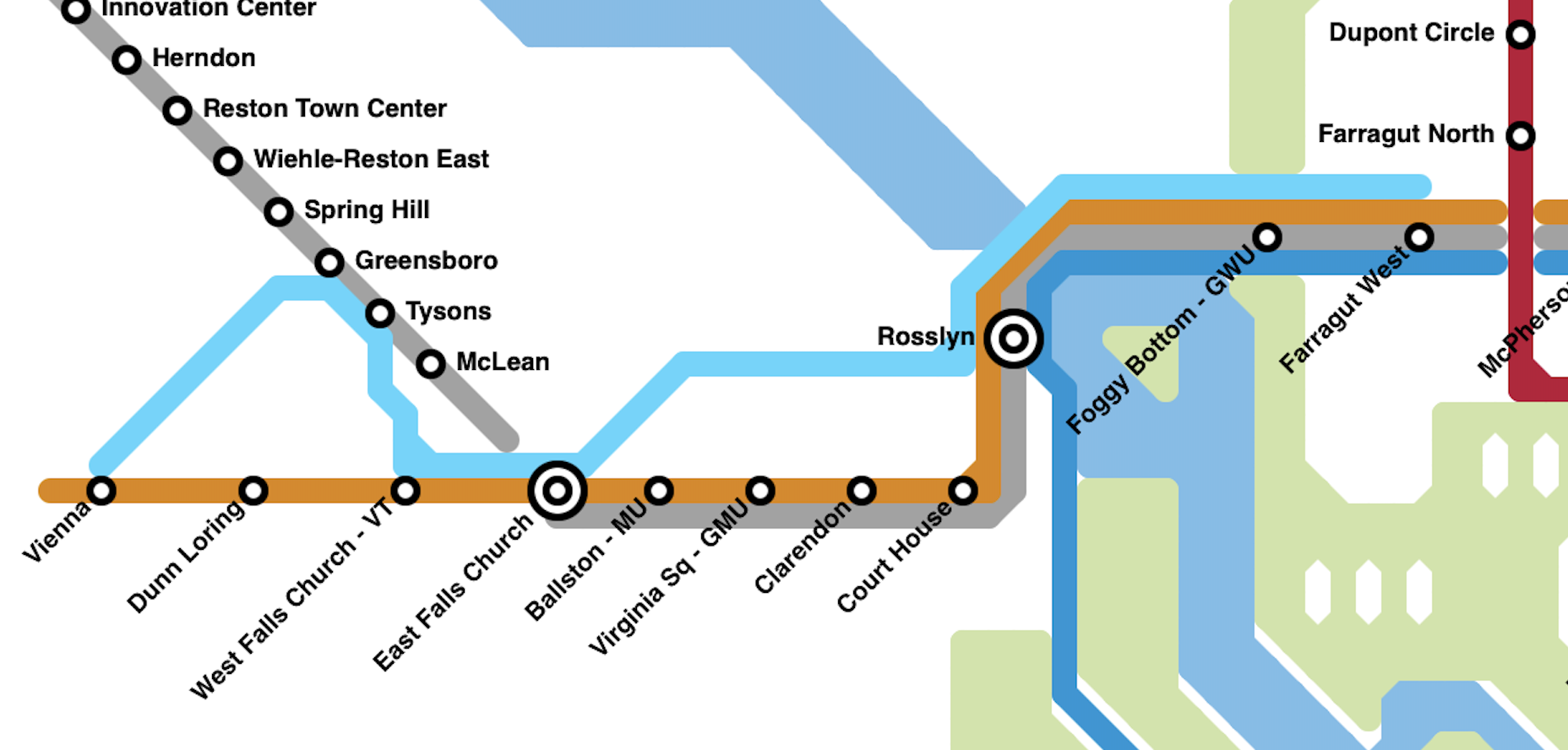}
    \caption{Map of WMATA metro with line in case 2 added in cyan}
    \label{fig:line2}
\end{figure}

\begin{figure}[H]
    \centering
    \includegraphics[width=0.5\linewidth]{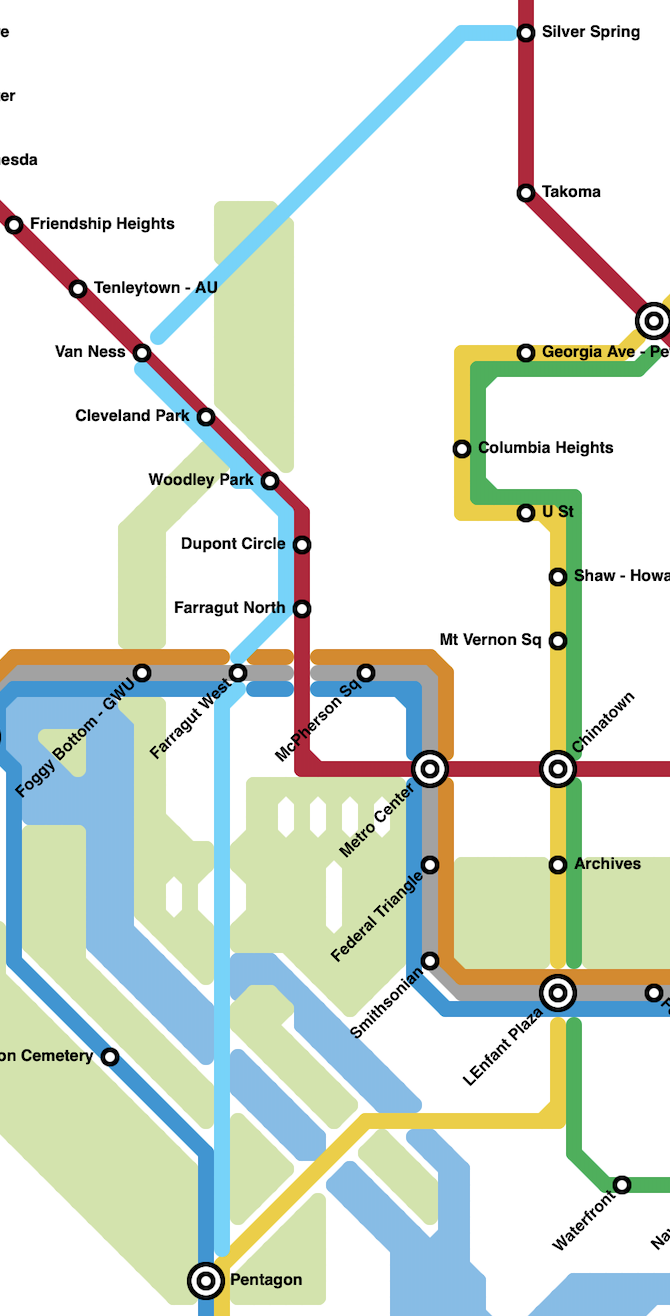}
    \caption{Map of WMATA metro with line in case 3 added in cyan}
    \label{fig:line3}
\end{figure}

\begin{figure}[H]
    \centering
    \includegraphics[width=0.5\linewidth]{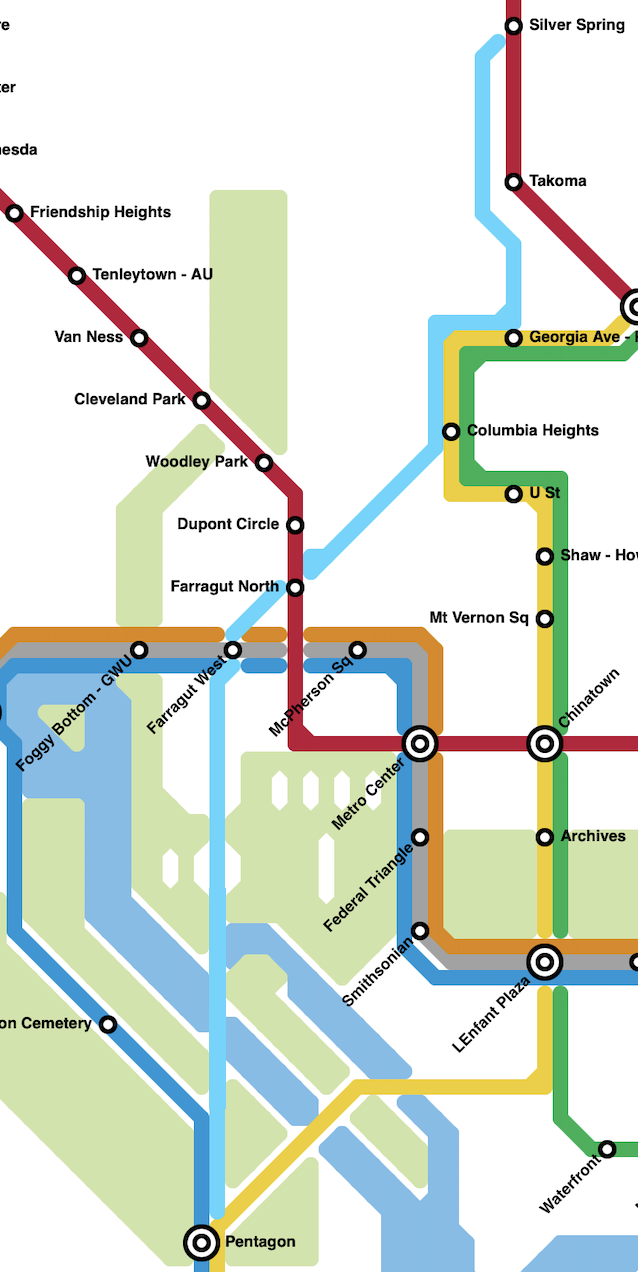}
    \caption{Map of WMATA metro with line in case 4 added in cyan}
    \label{fig:line4}
\end{figure}

\begin{figure}
    \centering
    \includegraphics[width=0.5\linewidth]{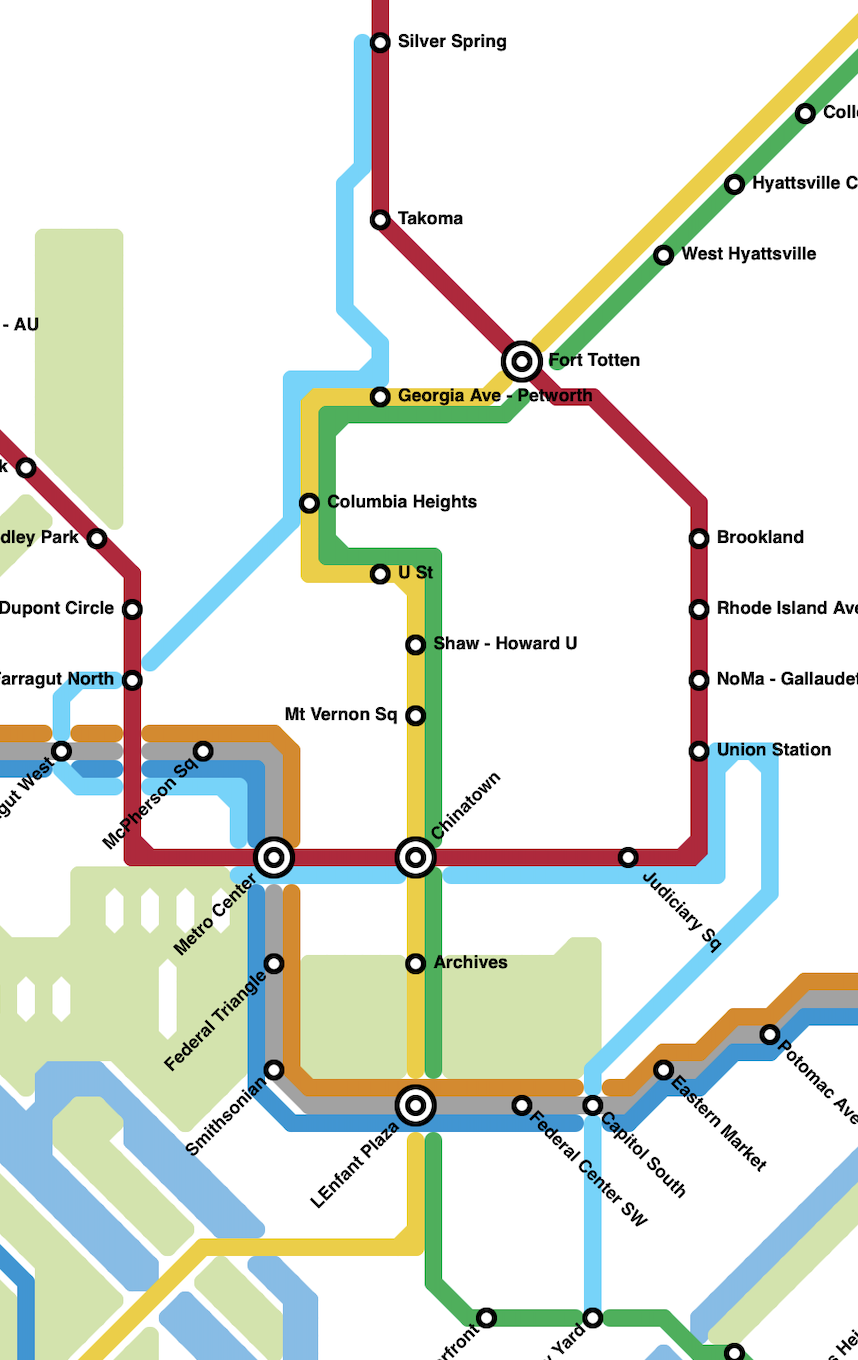}
    \caption{Map of WMATA metro with line in case 5 added in cyan}
    \label{fig:line5}
\end{figure}

\end{document}